\documentclass{aastex63}
\usepackage[utf8]{inputenc}

\usepackage{amsmath}
\usepackage{amssymb}

\shorttitle{Instabilities in Compact Super-Earth Systems}

\begin{document}

\title{Architectures of Compact Super-Earth Systems Shaped by Instabilities}
\accepted{February 27, 2022}
\shortauthors{Goldberg \& Batygin}

\correspondingauthor{Max Goldberg}
\email{mg@astro.caltech.edu}

\author[0000-0003-3868-3663]{Max Goldberg}
\affiliation{Department of Astronomy, California Institute of Technology\\1200 E. California Blvd\\Pasadena, CA 91125, USA}

\author[0000-0002-7094-7908]{Konstantin Batygin}
\affiliation{Division of Geological and Planetary Sciences, California Institute of Technology\\1200 E. California Blvd\\Pasadena, CA 91125, USA}

\begin{abstract}
Compact non-resonant systems of sub-Jovian planets are the most common outcome of the planet formation process. Despite exhibiting broad overall diversity, these planets also display dramatic signatures of intra-system uniformity in their masses, radii, and orbital spacings. Although the details of their formation and early evolution are poorly known, sub-Jovian planets are expected to emerge from their natal nebulae as multi-resonant chains, owing to planet-disk interactions. Within the context of this scenario, the architectures of observed exoplanet systems can be broadly replicated if resonances are disrupted through post-nebular dynamical instabilities. Here, we generate an ad-hoc sample of resonant chains and use a suite of N-body simulations to show that instabilities can not only reproduce the observed period ratio distribution, but that the resulting collisions also modify the mass uniformity in a way that is consistent with the data. Furthermore, we demonstrate that primordial mass uniformity, motivated by the sample of resonant chains coupled with dynamical sculpting, naturally generates uniformity in orbital period spacing similar to what is observed. Finally, we find that almost all collisions lead to perfect mergers, but some form of post-instability damping is likely needed to fully account for the present-day dynamically cold architectures of sub-Jovian exoplanets.
\end{abstract}

\section{Introduction}
Over the course of the past two decades, the discovery and characterization of thousands of extrasolar planets by the \textit{Kepler} and \textit{TESS} missions has shown that planet formation is both highly efficient and suggested that the dominant mode of planet formation is one that produces so-called super-Earths. These planets tend to exist in multiples, and typically have masses a few times that of Earth and orbital periods smaller than $\sim100$ days \citep{Howard2012, Batalha2013, Fressin2013, Marcy2014, Thompson2018}.
A remarkable discovery of this expanding census is the physical diversity of the galactic planet sample. Planets vary by several orders of magnitude in radius, mass, and orbital distance and frequently orbit stars not similar to the Sun \citep{Raymond2020}. While not fully quantified, the emerging picture suggests the Solar System is an unusual outcome of planet formation because of the presence of Jupiter and lack of a compact system of inner planets \citep{Batygin2015a, Izidoro2015, Raymond2018a}. 

An equally remarkable, but more recent discovery, is that the galactic diversity largely disappears when considering only individual planetary systems. The ``peas-in-a-pod'' pattern of intra-system uniformity has demonstrated that the dispersion in planet spacing, mass, and radius within individual planetary systems is much smaller than that across the exoplanet population as a whole \citep{Weiss2018, Millholland2017, Wang2017}. In other words, many systems seem to have a characteristic planet mass, radius, and spacing that is representative for a particular star, but differs drastically system-to-system. The physical origin of this uniformity remains unresolved.

A distinct mystery is the origin of the period ratio distribution. Plausible models of super-Earth formation typically include planet-disk interactions that drive inward migration and often lead to capture of the planets into mean-motion resonances (MMRs)---orbital configurations where the period ratios are approximated by nearby integers. \citep{Terquem2007}. While there is weak clustering of planet just wide of mean-motion resonances, near-resonant planets form a distinct minority in the close-in planet sample \citep{Fabrycky2014}. A rare, but important exception to this rule is the class of resonant chains, such as Kepler-60, Kepler-80, Kepler-223, K2-138, TRAPPIST-1, and TOI-178 \citep{Gozdziewski2016,MacDonald2016,Mills2016,Christiansen2018,Luger2017,Leleu2021}, as well as a subset of near-resonant systems that show hints of past resonant behavior \citep{Pichierri2019,Goldberg2021}. Nevertheless, the dominantly non-resonant orbital configurations of short-period planets constitute an important point of tension between theory and observations.

Multiple ideas have been put forth to explain this discrepancy over the last decade and a half. One suggestion is that disk turbulence destabilizes resonances for small planets \citep{Adams2008, Rein2009, Batygin2017}. However, both analytic calculations \citep{Batygin2017} as well as numerical simulations have confirmed that this effect is too small to explain the discrepancy \citep{Izidoro2017}. Likewise, resonant metastability, proposed in \cite{Goldreich2014}, operates in region of parameter space that does not encompass most of the sample. As a whole, these models have failed to provide a complete explanation for the data, and detailed hydrodynamic simulations \citep[e.g.,][and the references therein]{Cresswell2008, Ataiee2020} find that formation of compact resonant chains is a common outcome of disk-driven orbital evolution. Given this tension between theoretical expectations and observational ground-truths, physical processes must either prevent the formation of resonances in the first place, or disrupt them later.

The recently-proposed ``breaking the chains'' scenario of \citep{Izidoro2017} argues for the latter alternative. In this model, resonances \textit{are} in fact routinely established in nascent exoplanetary systems during orbital migration. Subsequently, the gaseous disk, which had provided eccentricity damping, dissipates, and the planetary system relaxes through the onset of dynamical instability and collisions. Several aspects of the exoplanetary sample are consistent with this hypothesis. First, planetary systems lie close to the margin of stability on Gyr timescales, suggesting that they experienced dynamical sculpting, i.e. encountering instabilities until becoming stable \citep{Pu2015}. Second, widespread instabilities reproduce the shape and slope of the observed period-ratio distribution if $\sim90\%$ of systems experience such a disruption in their lifetime \citep{Izidoro2017,Izidoro2021}. As successful as this scenario is in explaining many constraints of the observed planetary sample, an important outstanding problem remains. Naively, consolidation of planets during collisions could destroy the delicate intra-system uniformity that is observed. Furthermore, orbital eccentricities are excited by planet-planet scattering, but damped by collisions \citep{Matsumoto2017, Esteves2020, Poon2020} and it remains unclear whether measured low eccentricities of planets in compact systems are consistent with typical post-instability orbits \citep{Hadden2014, Mills2019, Yee2021}.



The remainder of this paper details our investigation into the compatibility of the observed peas-in-a-pod correlations with the instability model. We create physically motivated models of pre-instability super-Earth/sub-Neptune systems, trigger instabilities, and compute statistical properties of their post-instability architectures. We then compare them to observed results. In section 2, we describe how we construct physically realistic initial conditions of resonant systems informed by real, stable, resonant chain systems. In section 3, we describe how we trigger dynamical instabilities and track evolution of the systems through collisions and mergers. In section 4 we present the results of our suite of simulations and their degree of consistency with observed data. We summarize and discuss our results in section 5.

\section{Initial Conditions}
The first step in modeling the instability scenario is to construct a broad library of initial conditions. In the framework of this model, the observed resonant chains are the small fraction of systems that did not undergo episodes of post-nebular planet-planet scattering. Therefore, we construct our initial conditions informed by this observed subsample. By virtue of being near MMR, these systems lend themselves to precise mass determinations through transit timing variations \citep{Lithwick2012a}, and are well-studied with spectroscopic surveys \citep[e.g.][]{Petigura2017}.

\begin{deluxetable*}{cccccc}
\tablehead{\colhead{System} & \colhead{\# planets} & \colhead{$\overline{m}/(M_\star/M_\odot)$ $(\text{M}_\oplus)$} & \colhead{$\sigma_m/\overline{m}$} & \colhead{Resonances present} & \colhead{Source of mass measurements}}
\tablecaption{Basic properties of the six well-characterized resonant chains with sub-Jovian planets.
\label{tab:chains}}
\startdata
Kepler-60 & 3 & 3.91 & 0.18 & 4:3, 5:4 & \cite{Jontof-Hutter2016} \\
TRAPPIST-1 & 7 & 11.51 & 0.46 & 5:3, 8:5, 3:2, 4:3 & \cite{Agol2021} \\
Kepler-223 & 4 & 5.63 & 0.31 & 3:2, 4:3 & \cite{Mills2016} \\
Kepler-80 & 6\tablenotemark{a}\tablenotemark{b} & 8.43 & 0.23 & 3:2, 4:3 & \cite{MacDonald2016} \\
TOI-178 & 6 & 6.41 & 0.53 & 2:1, 5:3, 3:2 & \cite{Leleu2021} \\
K2-138 & 5\tablenotemark{b} & 7.06 & 0.62 & 3:2 & \cite{Christiansen2018} \\
\enddata
\tablenotetext{a}{Only 4 planets in Kepler-80 have measured masses}
\tablenotetext{b}{Kepler-80 f and K2-138 g have been excluded because they are decoupled from the resonant dynamics}
\end{deluxetable*}

\begin{figure}
    \centering
    \includegraphics[width=0.7\textwidth]{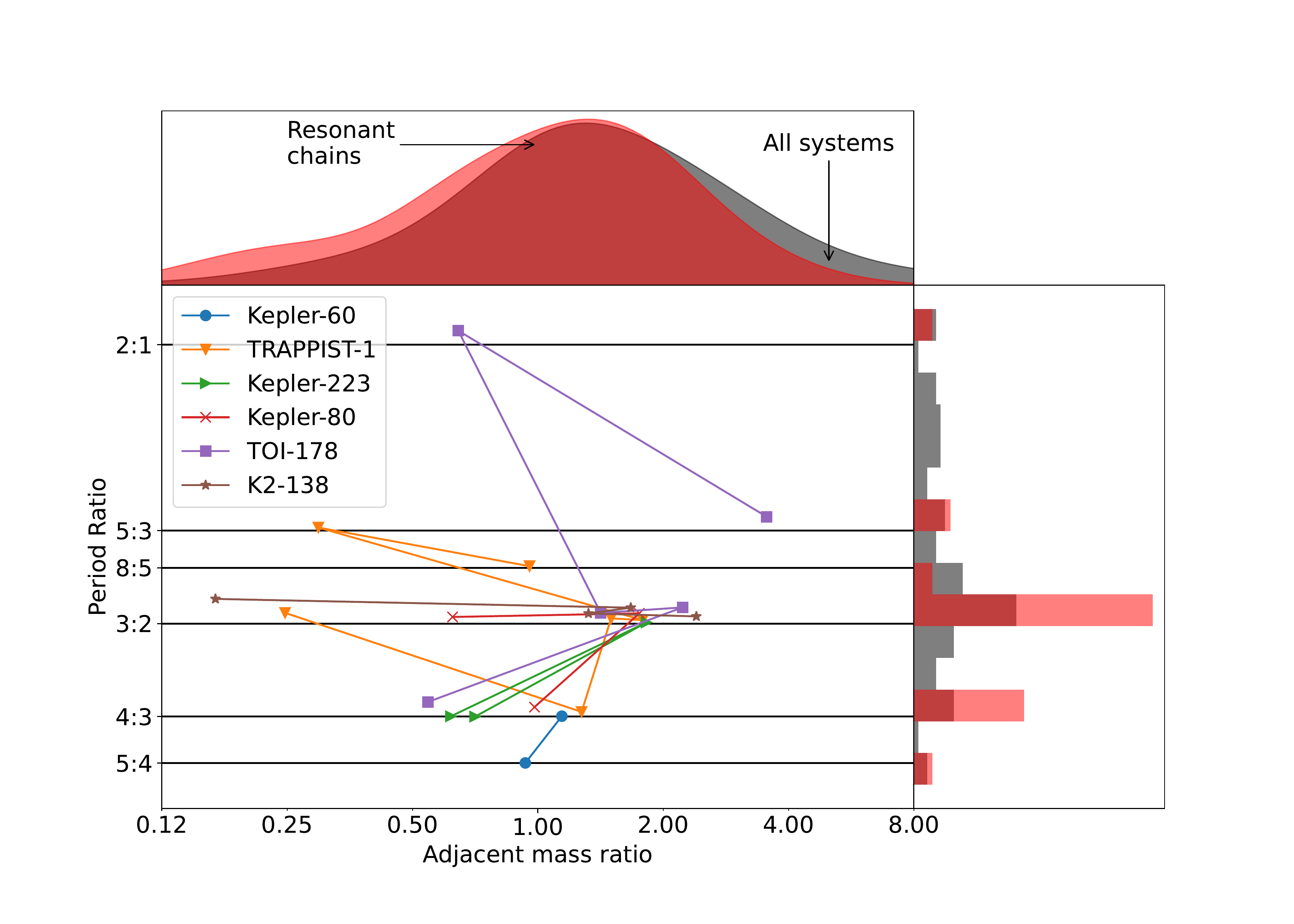}
    \caption{Mass ratios and orbital period ratios of five well-characterized resonant chains. Center: colored points represent adjacent pairs of planets and are placed according to mass and orbital period ratio computed from outside-in. Colored lines connect adjacent pairs in the same system, and horizontal black lines mark orbital resonances. Adjacent pairs missing at least one measured mass were discarded. Top: kernel density estimates of the distribution of mass ratio for the five resonant chains (red) versus all sub-Jovian systems (gray). Right: histogram of period ratios of adjacent planets with the same color scheme as the top panel. Overall, resonant chains exhibit tighter clustering in both period ratios and mass ratios than the overall sample of sub-Jovian exoplanets.}
    \label{fig:chains}
\end{figure}

To construct multi-resonant systems through convergent orbital migration, we select the number of planets $N$, average planet mass $\overline{m}$, planet relative standard deviation $\sigma_m/\overline{m}$, and initial resonant indices $p:q$. We take $N = 11$, approximately twice the inferred true average multiplicity \citep{Zink2019}. To cover a similar distribution of masses as the observed resonant chains (Table \ref{tab:chains}), we pick values of $\overline{m}=10.0$ for a higher-mass sample (runs 1-8) and $\overline{m}=1.5$ for a lower-mass sample (run 9), typically selecting the highest mass for which the resonances can be formed without triggering an instability in the presence of the disk. Initial mass dispersions $\sigma_m/\overline{m}$ are in the range $0$ to $0.5$. Each simulation draws masses from a normal distribution with mean $\overline{m}$ and variance $\sigma_m^2$. While real resonant chain systems, such as the ones in Fig \ref{fig:chains}, contain a variety of first-, second-, and third-order resonances, our constructed systems must be more compact than the observed resonant systems in order to go unstable. Therefore, we pick resonances with smaller period ratios, specifically 4:3 for the high-mass sample and 5:4 for the low-mass sample. These initial conditions are summarized in Table \ref{tab:sims}.

We simulate resonant chain formation and subsequent evolution with the \texttt{mercurius} integrator from the \texttt{rebound} gravitational dynamics software package, with timesteps lower than 1/15 the innermost orbital period \citep{Rein2019}. Planets are placed on circular, coplanar orbits around an $M_\star=1M_\odot$ star, with the semi-major axis of the innermost planet set to $0.1$ AU and period ratios just wide of the intended resonance. We then activate convergent migration with eccentricity damping, implemented within \texttt{reboundx} \citep{Tamayo2020}, until the planets have entered the intended resonance ($\lesssim 10^5$ yr). The details of the migration and damping timescales are provided in Appendix \ref{app}. In practice, planets in the lower-mass simulations entered either the 5:4 or the 6:5 resonance. Then, we remove semi-major axis and eccentricity damping exponentially over a timescale of $10^3$ yr. We have checked that increasing these timescales does not meaningfully alter our results. The final pre-instability systems contain many librating resonance angles and the orbital orbital eccentricities are typically $\sim 0.05$ or smaller. At this point we rescale the systems so that the inner planet has semi-major axis 0.1AU, which corresponds to a drop-off in prevalence of super-Earths \citep{Petigura2013}. 

\section{Instabilities}
The instability and collision-driven model necessitates a source of instabilities. To this end, \cite{Izidoro2017} and \cite{Izidoro2021} produce post-disk-dissipation planetary systems that are too tightly packed to remain stable on $\sim$ Gyr timescale, and hence will undergo an intrinsic dynamical instability triggered purely by gravitational dynamics. However, there are many other possible instability mechanisms due to extrinsic, i.e. astrophysical, factors. \cite{Spalding2016} and \cite{Spalding2018} demonstrate that an oblique and initially rapidly-rotating star can excite mutual inclinations, leading to secular resonances that drive instabilities \citep[see also][]{Schultz2021}. \cite{Matsumoto2020} show that mass loss in the systems (of order $\sim 10\%$ in planetary mass or $\sim 1\%$ in stellar mass) can also induce instabilities and break resonant chains. As a whole, if instabilities occur frequently, they unavoidably play a major role in modifying orbits and shaping the architectures of exoplanetary systems \citep{Ogihara2009}.

\begin{figure}
    \centering
    \includegraphics[width=0.9\textwidth]{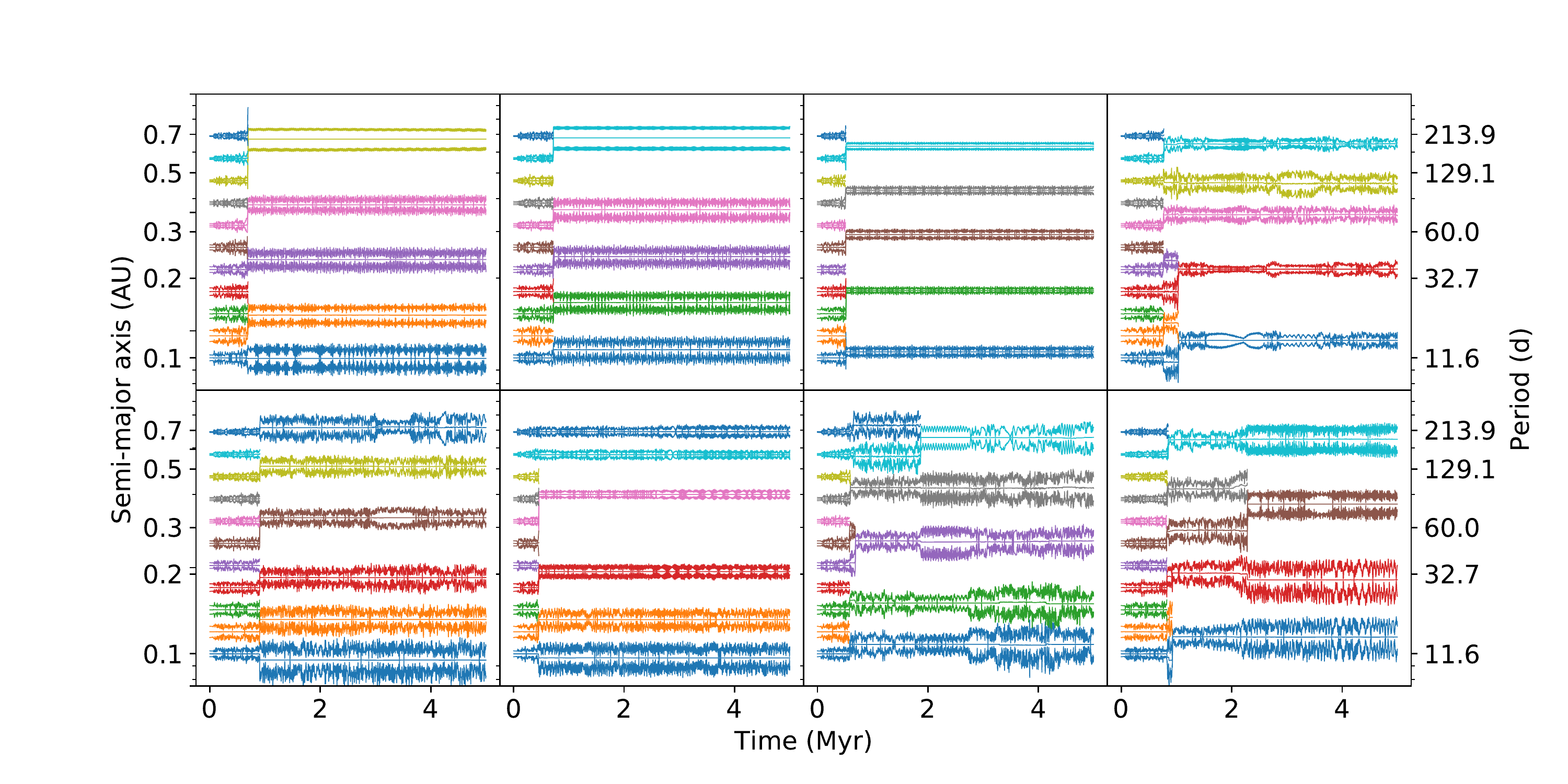}
    \caption{Eight collisional outcomes for the same initial system with $N=11$, $\overline{m}=8.6M_\oplus$, initial resonance 4:3, and initial mass dispersion $\mathcal{D}\approx 0.1$. For each planet, the semi-major axis $a$, pericenter distance $a(1-e)$, and apocenter distance $a(1+e)$ are plotted in the same color. When two planets collide, the traces retain the color of the inner planet. Instability-driven evolution leads to wider orbital spacing, while retaining a degree of mass-uniformity that is compatible with the data.}
    \label{fig:unstable_a}
\end{figure}

Our intention is not to test various instability mechanisms; rather, we want to investigate the consequences of collisions and mergers. Additionally, the instability mechanism is not believed to dramatically affect the post-instability configuration itself \citep{Nesvorny2012}. Therefore, we adopt the mechanism of \cite{Matsumoto2020}: planet masses are exponentially decreased with an evolution timescale of 1 Myr until they reach 90\% of their original mass. This suffices to trigger instabilities in many cases without qualitatively changing the system and does not require overly long integrations. We evolve the initially resonant systems for a further 5 Myr monitoring for collisions. When one is detected, we record the colliding pair's masses and relative velocities and then replace them with a single planet whose mass and linear momentum are the sum of the colliders'. While this assumes collisions are perfect mergers, we verify this assumption below, in agreement with the results of \cite{Poon2020} and \cite{Esteves2022}. To produce a statistically useful sample, we repeat the collision phase 50 times, starting from the mass reduction, but use a mass loss timescale that is randomly shifted by $\sim 1\%$ from 1 Myr. Because of the chaotic dynamics of planet-planet scattering, each run produces a different set of collisions and it is possible to compute distributions of final parameters (Figure \ref{fig:unstable_a}).

\section{Results}

To evaluate whether collisions are consistent with the architecture of observed planetary systems, we compute statistical measures used in previous works to characterize the mass and spacing uniformity of our synthetic systems. To construct the sample of observed systems with which to compare our synthetic ones, we select all systems from the Exoplanet Archive\footnote{\url{exoplanetarchive.ipac.caltech.edu}} with at least 3 planets that do not contain planets more massive than $30 \text{M}_\oplus$. The latter constraint is chosen because mass uniformity vanishes in systems with giant planets \citep{Wang2017}. The six resonant chains in Table \ref{tab:chains} and Figure \ref{fig:chains} are a subset of this sample.

\subsection{Intra-system Uniformity}
Although the works that identified the intra-system uniformity pattern each used different statistics to identify the uniformity \citep{Millholland2017,Wang2017,Weiss2018}, for definitiveness, we adopt a modified version of the mass uniformity metrics from \cite{Wang2017}. Specifically, we normalize by the average planet mass in a system, so that larger planets do not appear less uniform, and by the total number of systems, so that the metric does not depend on the number of systems. Hence, we define
\begin{equation}
            \mathcal{D} = \frac{1}{N_\text{sys}} \sum_{i=1}^{N_\text{sys}}\sqrt{\frac{\sum_{j=1}^{N_\text{pl}} (m_j - \overline{m}_i)^2}{\overline{m}^2_i(N_\text{pl} - 1)}} = \frac{1}{N_\text{sys}} \sum_{i=1}^{N_\text{sys}} \frac{\sigma_m}{\overline{m}_i}.
\end{equation}
Here, the inner sum is taken over a single system: $N_\text{pl}$ is the number of planets in a system and $m_j$ is the individual mass of the $j$-th planet. The outer sum is taken over all systems: $N_\text{sys}$ is the total number of systems considered and $\overline{m}_i=\sum_{j=1}^{N_\text{pl}} m_j/N_\text{pl}$ is the average planet mass in a system. The metric $\mathcal{D}$ is dimensionless, and can be understood as the average relative standard deviation in mass. It is also closely related to the mass partitioning $Q$ defined in \cite{Gilbert2020}, differing by a square root and a factor of $N_\text{pl}$. A similar expression for uniformity in radius can be defined, but we do not explicitly use it because radii in our simulations are computed directly from the mass and assume a constant density. The uniformity metric for our multiplanet sample is $\mathcal{D} = 0.48$, and $\mathcal{D}=0.37$ for the six resonant chains. Hence, the resonant chains are somewhat more uniform in mass than the full population (see also Figure \ref{fig:chains}).
\begin{figure}
    \centering
    \includegraphics[width=0.5\textwidth]{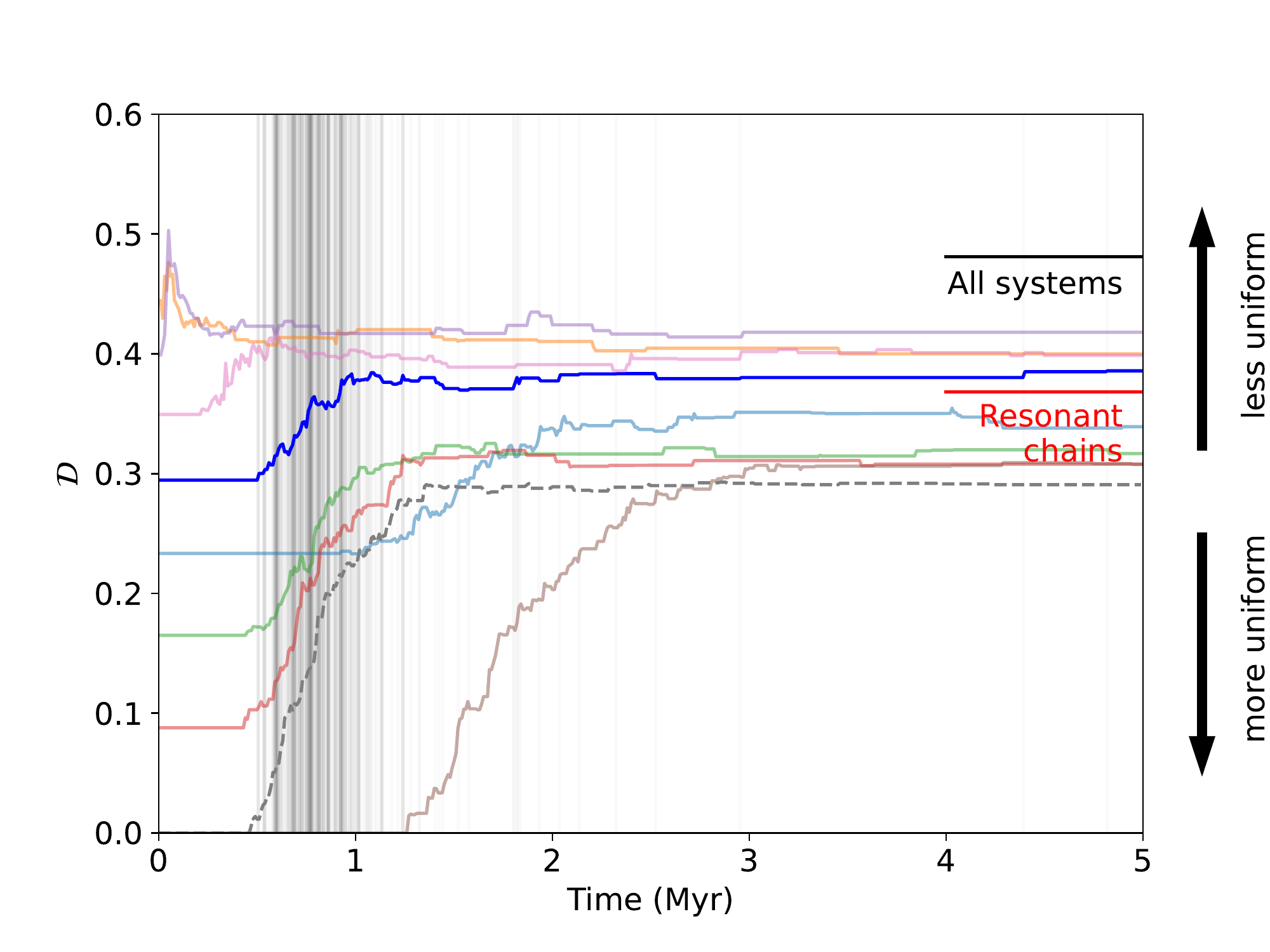}
    \caption{Runs of 50 integrations of nearly the same initial conditions, with $N=11$, $\overline{m}\sim 10M_\oplus$, and initial resonance 4:3, but differing initial mass dispersion $\mathcal{D}$. The dashed line marks the evolution for run 9, which has different initial conditions (see Table \ref{tab:sims}). The red and black horizontal lines represent $\mathcal{D}$ for the resonant chains from Section 2 and all systems, respectively. For the run starting around $\mathcal{D}\approx 0.35$, plotted darker in blue, translucent vertical lines indicate the time of a collision, which triggers a change in $\mathcal{D}$.}
    \label{fig:D}
\end{figure}

As collisions combine planets, $\mathcal{D}$ varies significantly, and its final value depends strongly on which planets collide. Accordingly, we take the average of $\mathcal{D}$ across the 50 runs and track it as a function of time. The evolution of the average $\mathcal{D}$ for the high-mass sample is shown in Figure \ref{fig:D}. During the 5 Myr integration, the number and masses of planets change as planets collide and merge or, rarely, are ejected from the system. For more uniform initial conditions $\mathcal{D}$ generally increases, and settles to a value of $\sim 0.3-0.4$. This lies slightly above the observed $\mathcal{D}$ of 0.37 for resonant chains and below 0.48 for all systems with $M_\text{max} < 30\text{M}_\oplus$, meaning that even after dynamical relaxation and the associated collisions, this set of post-instability systems is marginally more uniform than the overall Kepler sample.

Surprisingly, the final value of $\mathcal{D}$ does not strongly depend on the initial mass distribution. We ran 8 suites of simulations with $\overline{m}\sim10\text{M}_\oplus$, 4:3 resonances, and initial $\mathcal{D}$ varying from 0 to 0.45. The results, shown as translucent lines in Figure \ref{fig:D}, indicate that the cascade of collisions does not necessarily increase $\mathcal{D}$, but brings it within a range $0.3 - 0.4$. This suggests that an arbitrary choice of initial $\mathcal{D}$ does not significantly bias the results. 

\subsection{Hill spacing and period uniformity}
A straightforward consequence of an instability phase is that the post-instability system must be stable on $\sim$ Gyr timescales. This manifests as an increase in the Hill spacing. First, we define the mutual Hill radius as
\begin{equation}
    R_{Hj} = \left(\frac{m_{j+1}+m_j}{3M_\star}\right)^{1/3} \frac{a_{j+1} + a_j}{2},
    \label{eq:RH}
\end{equation}
which represents a characteristic length scale for gravitational interactions between planets. Then, the Hill spacing is 
\begin{equation}
    \Delta_j = \frac{a_{j+1} - a_j}{R_{Hj}},
    \label{eq:Delta}
\end{equation}
and the average Hill spacing $\overline{\Delta}$ is simply the average of $\Delta_j$ in a system. The lifetime of a multiplanet system strongly depends on $\overline{\Delta}$ \citep{Chambers1996}, so collisions will proceed until $\overline{\Delta}$ grows and the system relaxes. The final values of $\overline{\Delta}$ in Table \ref{tab:sims} are $10-13$ for the high-mass sample, comparable to observed compact multiplanet systems \citep{Pu2015}.

\begin{deluxetable*}{ccccccccc}
\tablehead{\colhead{Run} & \colhead{initial $\overline{m}$ $(\text{M}_\oplus)$} & \colhead{final $\overline{m}$ $(\text{M}_\oplus)$} & initial resonance & \colhead{initial $\mathcal{D}$} & \colhead{final $\mathcal{D}$} & \colhead{final $r$} & \colhead{final $\overline{\Delta}$} & \colhead{final $f$}}
\tablewidth{\textwidth}
\tablecaption{Key system architecture parameters in our suite of simulations with $N=11$.
\label{tab:sims}}
\startdata
1 & 8.0 & 16.6 & 4:3 & 0.00 & 0.31 & 0.34 & 10.2 & 0.46 \\
2 & 8.6 & 17.1 & 4:3 & 0.09 & 0.31 & 0.44 & 11.6 & 0.43 \\
3 & 9.1 & 18.2 & 4:3 & 0.16 & 0.32 & 0.57 & 11.4 & 0.43 \\
4 & 9.6 & 20.2 & 4:3 & 0.23 & 0.34 & 0.36 & 10.1 & 0.46 \\
5 & 10.2 & 21.9 & 4:3 & 0.29 & 0.39 & 0.46 & 11.4 & 0.41 \\
6 & 10.7 & 23.7 & 4:3 & 0.35 & 0.41 & 0.53 & 12.1 & 0.43 \\
7 & 11.3 & 24.7 & 4:3 & 0.40 & 0.42 & 0.36 & 12.8 & 0.39 \\
8 & 11.8 & 26.5 & 4:3 & 0.44 & 0.40 & 0.46 & 13.1 & 0.37 \\
9 & 1.50 & 2.64 & 5:4, 6:5 & 0.00 & 0.30 & 0.44 & 16.5 & 0.46 \\
\enddata
\end{deluxetable*}

\begin{figure}
    \centering
    \includegraphics[width=0.9\textwidth]{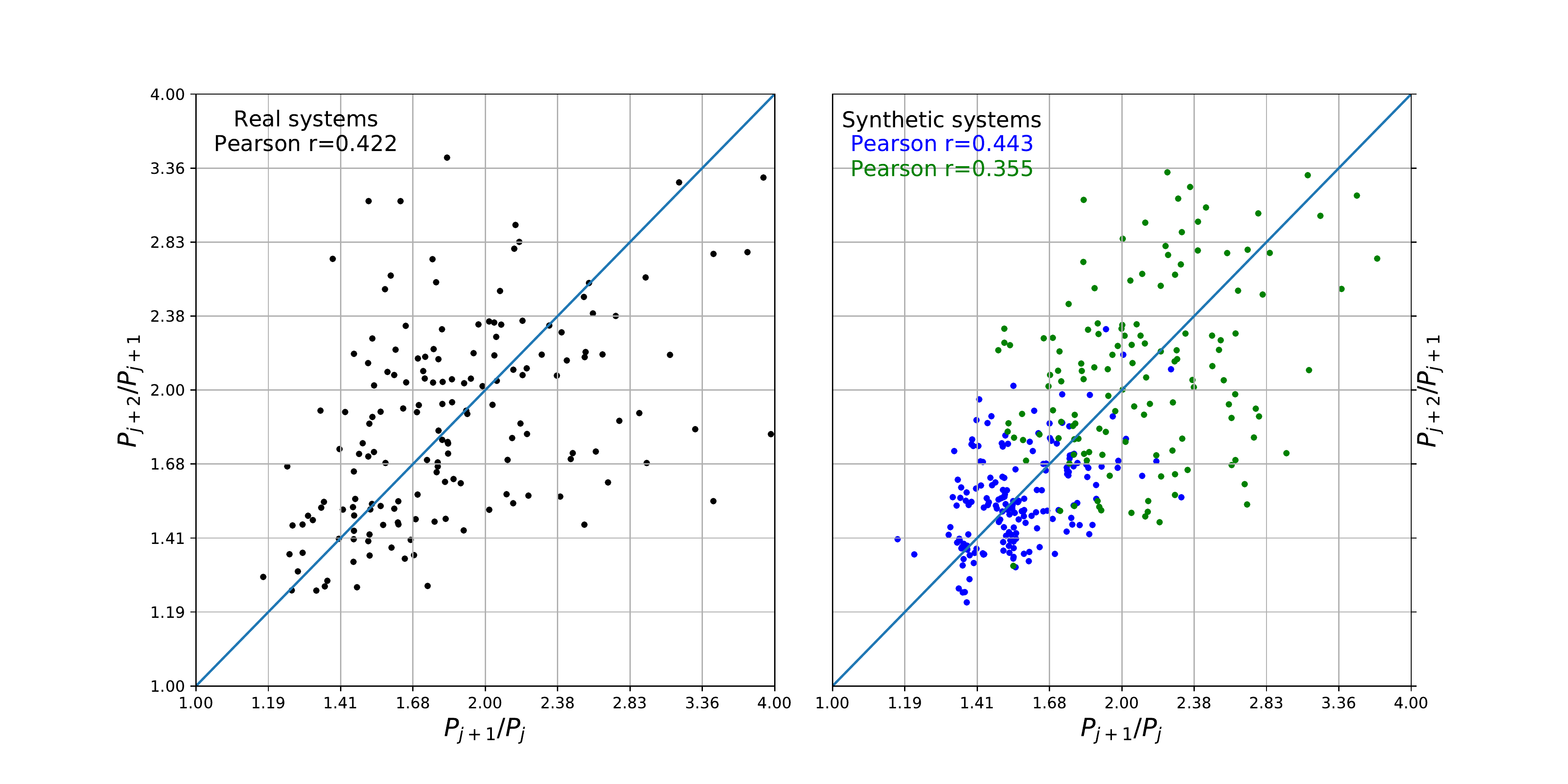}
    \caption{Period ratios of outer planet pairs and inner planet pairs for the well-characterized systems in \cite{Weiss2018} (left) and synthetic systems (right). In the right panel, blue and green dots correspond to the low-mass and high-mass samples, respectively. The Pearson $r$ correlation between the orbital period ratios is statistically significant ($p<10^{-4}$) in both cases.}
    \label{fig:per_corr}
\end{figure}

Because Equations \ref{eq:RH} and \ref{eq:Delta} depend exclusively on planet mass and semi-major axis, intra-system uniformity in mass and Hill spacing directly implies uniformity in semi-major axis ratio and therefore period spacing. To quantify this, we adopt the metric from \cite{Weiss2018}, which is to compute the Pearson $r$ correlation coefficient of period ratios of adjacent pairs of planets, i.e. $P_{i+2}/P_{i+1}$ and $P_{i+1}/P_i$. With their sample of well-characterized planets, they find a correlation of 0.46 and high statistical significance. Our simulations broadly reproduce this in the high- and low-mass simulations this with an average correlation of $\overline{r}=0.44$ and some scatter (Table \ref{tab:sims}, Figure \ref{fig:per_corr}).

\subsection{Period ratio distribution}

\begin{figure}
    \centering
    \includegraphics[width=0.5\textwidth]{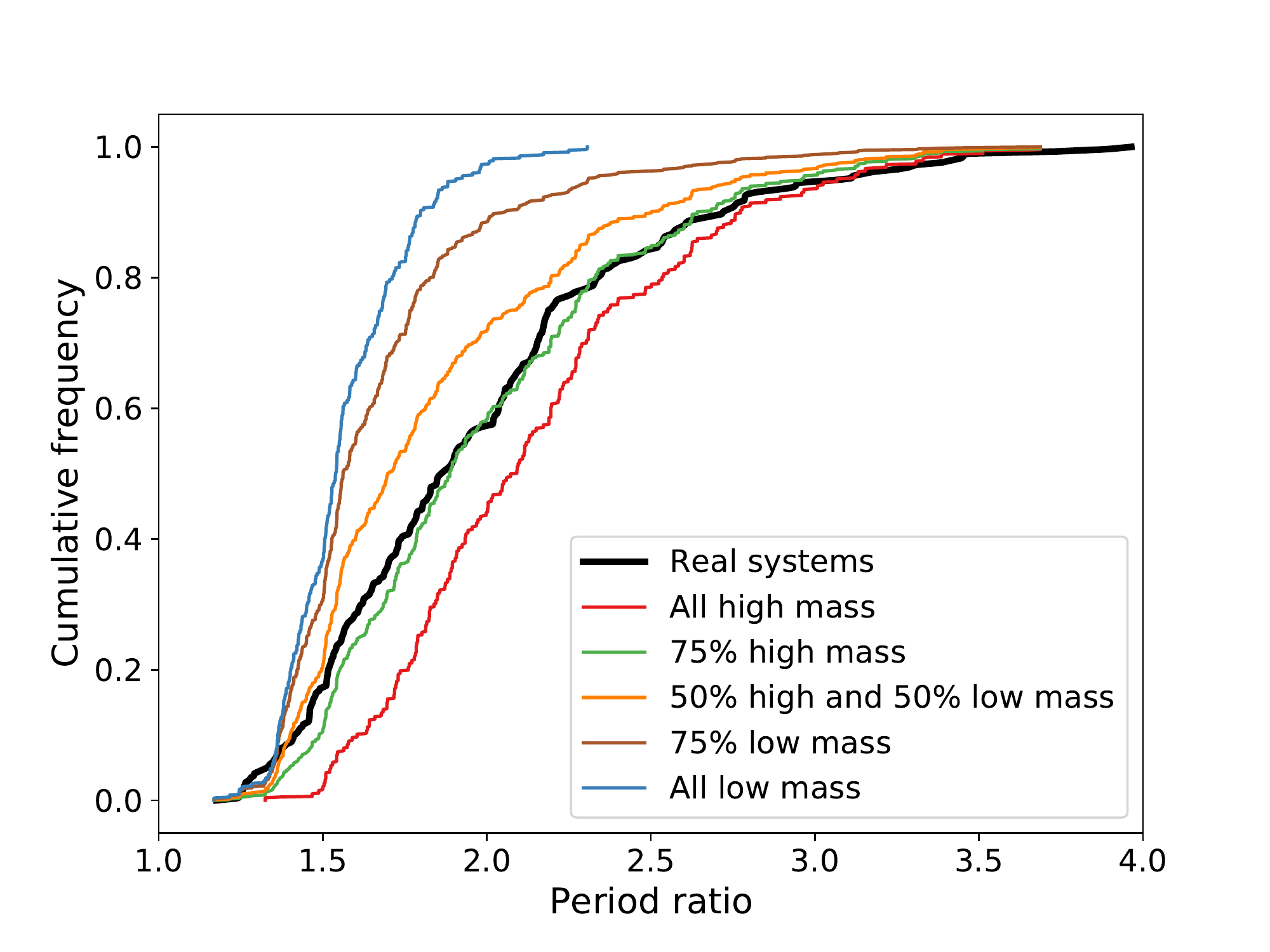}
    \caption{Cumulative frequency of period ratio for the post-instability synthetic systems (colored lines) and real systems (black line). Blended samples are constructed by mixing the two populations of low- and high-mass planets in the listed fractions.}
    \label{fig:rat_hist}
\end{figure}

The principal difference between our high-mass and low-mass systems is the period ratio distribution. Higher-mass systems must be more widely spaced to ensure stability. The high-mass sample lacks almost any planet pairs with period ratio less than 1.5 but otherwise matches the slope of the cumulative distribution. On the other hand, the low-mass sample misses period ratios above 2.0. This suggests that we should combine the samples in a particular proportion to produce an optimal match to the data. A similar approach was taken in \cite{Izidoro2017,Izidoro2021}. We create the blended populations by choosing a fraction of systems to draw from the low-mass sample (run 9) while drawing the remainder from run 5 of the high-mass sample, which has average mass and initial dispersion similar to the resonant chains. Figure \ref{fig:rat_hist} shows the results of this exercise. A mixture of $25\%$ of systems taken from the low-mass sample and $75\%$ from the high-mass one fits the period ratio distribution best. We emphasize that these numbers are not to be taken literally---super-Earth planetary systems do not form in these two discrete mass ranges---but we highlight that a simple model of two populations reproduces many aspects of the observed sample with surprising ease. Because the uniformity observed in super-Earth systems is confined to planets that orbit a common star, it is not suppressed by combining a diverse set of systems. Accordingly, this merged sample has a period ratio correlation of $r=0.56$ and intra-system mass dispersion $\mathcal{D}=0.33$.

\subsection{Collisions}
While our simulations treat impacts as perfect mergers, the outcomes of planetary-scale collisions in general depend strongly on the speed and angle of the encounter as well as the mass ratio of the colliders \citep{Stewart2012}. Consider a projectile of mass $m'$ and radius $s'$ that collides with a target of mass $m$ and radius $s$ at a relative velocity $V_\text{imp}$ and impact angle $\theta$. Only a fraction of the projectile interacts with the target, specifically the interacting mass
\begin{equation}
    m'_\text{interact} = \frac{3s'l^2 - l^3}{4s'^3} m'
\end{equation}
where $l$ is the projected length
\begin{equation}
    l = \left(s + s'\right)(1 - \sin{\theta}).
\end{equation}
\citep{Leinhardt2012}. Numerical simulations have shown that collisions are nearly perfect mergers if the collisional speed does not exceed the escape velocity of the newly formed planet,
\begin{equation}
    V_\text{esc} = \sqrt{2G(m+m'_\text{interact})/S}
    \label{eq:Vesc}
\end{equation}
where $G$ is the gravitational constant and $S=(s^3+s'^3)^{1/3}$ is the radius of the new planet, assuming constant density \citep{Stewart2012}. 

\begin{figure}
    \centering
    \includegraphics[width=0.8\textwidth]{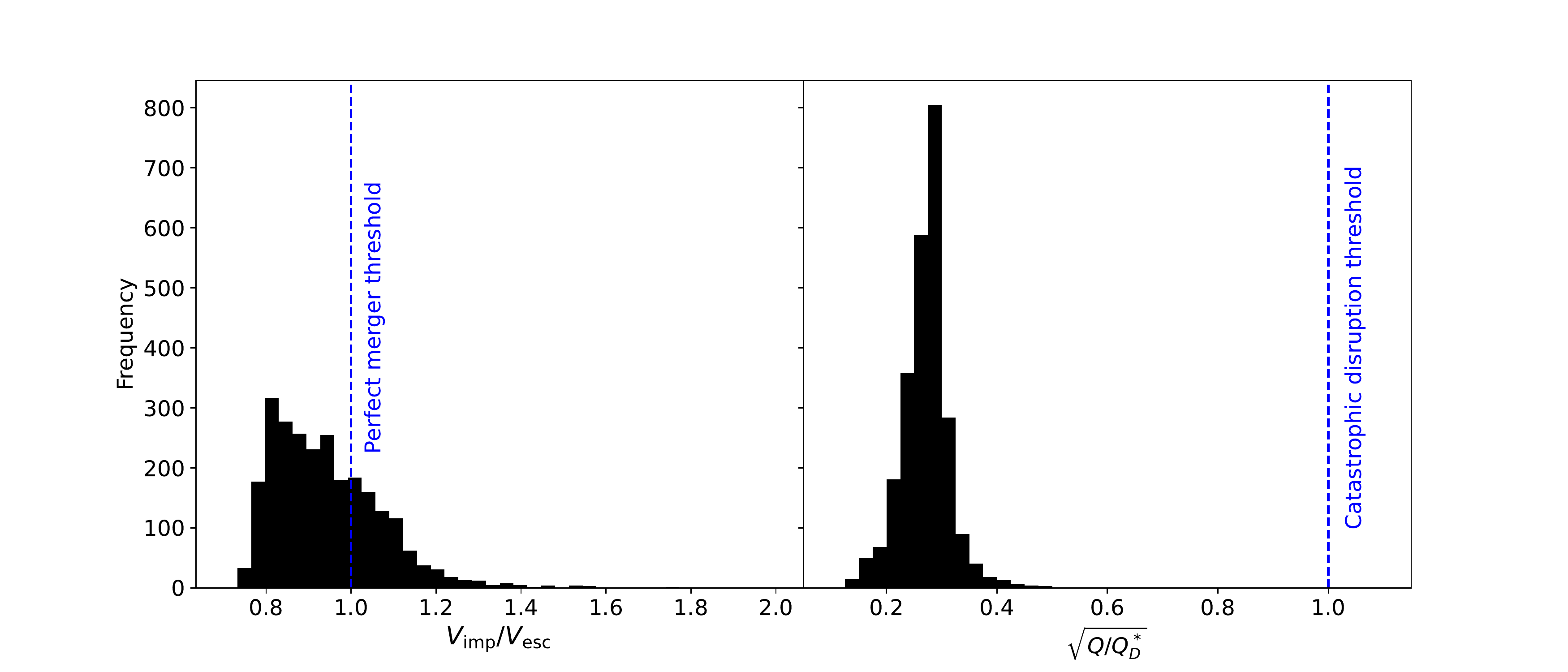}
    \caption{Collisional energetics of impacts in our simulations. Left panel: the distribution of the relative speed just before a collision, scaled by the escape velocity of the new planet. Collisions with $V_\text{imp}/V_\text{esc} < 1$ are expected to be perfect mergers. One event with $V_\text{imp}/V_\text{esc} \approx 12.4$ has been omitted for clarity. Right panel: the distribution of specific collision energy to the energy required for catastrophic disruption. The event omitted from the left panel lies just below $Q/Q_D^*=1$.}
    \label{fig:speed}
\end{figure}

The left panel of Figure \ref{fig:speed} shows the ratio of collision speed to escape velocity for the 2529 collisions in our simulations. While all collisions occur above $0.7V_\text{esc}$ due to mutual gravitation of the planets as they come together in the collision, approximately $70\%$ of collisions occur below the final escape velocity. Of the $\sim 30\%$ of collisions with $V_\text{imp} > V_\text{esc}$, most are just above the threshold for merging except for one collision with $V_\text{imp} \approx 12 V_\text{esc}$, not shown in the histogram. This unusual event likely resulted from a retrograde orbit formed during the scattering process.

Nevertheless, even collisions above the escape velocity do not necessarily disperse material completely. Specifically, for the projectile to catastrophically disrupt and unbind the target of mass into two or more pieces, the specific impact energy $Q$ must exceed the catastrophic disruption threshold $Q_D^*$, where
\begin{equation}
    Q = \frac{m' V_\text{imp}^2}{2(m+m')}
\end{equation}
and, in the gravity dominated regime,
\begin{equation}
    Q_D^* = q_g \rho_m \left(\frac{s}{1\text{cm}}\right)^b
\end{equation}
where, for high-speed collisions of basalt, $q_g \approx 0.5 \text{ erg cm}^3 \text{g}^{-2}$, $\rho_m\approx 3 \text{ g cm}^{-3}$ is the density, and $b=1.36$ \citep{Armitage2010}. The right panel of Figure \ref{fig:speed} shows the ratio $\sqrt{Q/Q_D^*}$ for the same collisions. All events lie below the catastrophic disruption threshold, including the exceptional event referred to above, which has $\sqrt{Q/Q_D^*}=0.98$.

These results are broadly consistent with those of \cite{Poon2020}. They use a different definition of escape velocity that is, in practice, always smaller than Equation \ref{eq:Vesc}, but nonetheless find that the majority of collisions occur only slightly above the escape velocity of the merged planet. They show furthermore that typical collisions do not dramatically change the ice fraction, but can strip gaseous envelopes. Similarly, \cite{Esteves2022} find that, while fragmentation during collisions can occur, the total amount of material that is stripped is small and has little effect on the dynamics.

\section{Discussion}
This work investigates the implications of dynamical instabilities and collisions on compact multiplanet system architectures. Within the context of this picture, we argue that the currently observed sub-sample of multi-resonant chains constitutes an adequate set of initial conditions for the instability model, and from this we conduct a suite of simulations to quantify the outcome of breaking the resonant locks. By and large, our calculations show that intra-system uniformity in mass, seen in resonant chains, is preserved after collisions and mergers in a way that is consistent with observations. Furthermore, as the planetary orbits are dynamically sculpted, a smooth period ratio distribution and period spacing uniformity naturally arise. Finally, we demonstrate that typical collisions are slow and unlikely to disrupt a large fraction of the planets.

An intriguing feature exhibited by the observational data is that the degree of orbital packing correlates inversely with the average planetary mass. That is to say, low-mass planets occupy more compact orbital architectures than their more massive counterparts (Weiss et al., in prep). This feature is distinct from a simple requirement of uniformity and long-term dynamical stability. For example, the Titius-Bode law is reflective of a period uniformity in the Solar System, despite a lack of mass uniformity \citep{Hayes1998}. The fact that a correlation which links mass and spacing exists hints that beyond any disk-driven processes that may regulate the terminal masses of forming planets \citep[e.g.][]{Lambrechts2014,Ormel2017}, the planetary masses themselves play a role in regulating the terminal spacings. Early dynamical evolution driven by transient instabilities provides the most natural mechanism to produce this feature in the data.


\begin{figure}
    \centering
    \includegraphics[width=0.9\textwidth]{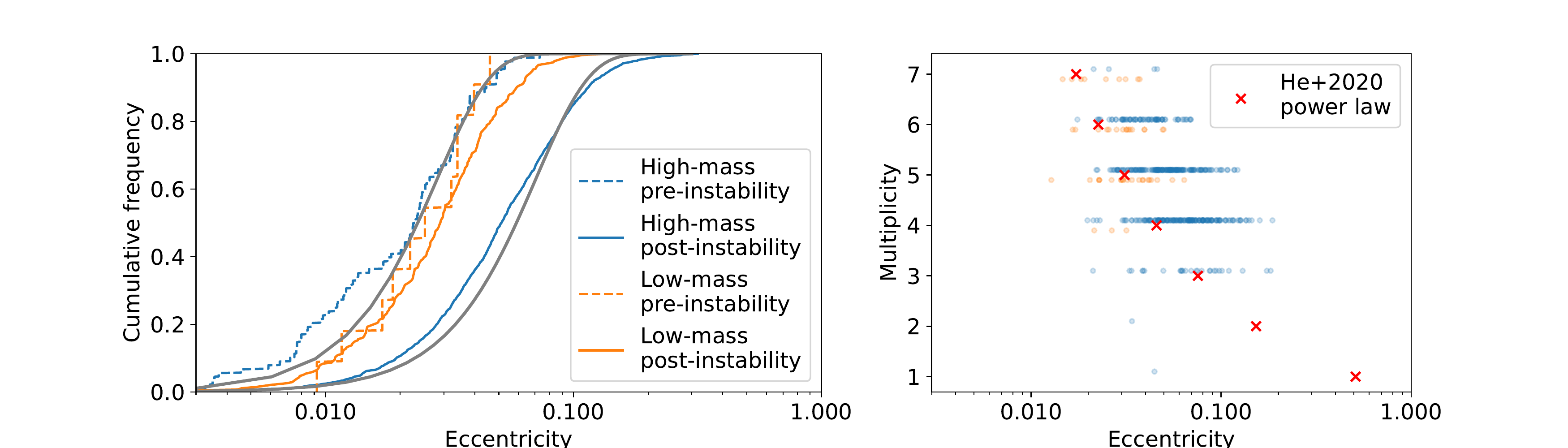}
    \caption{Eccentricity distributions for our suite of simulations. Left panel: Cumulative frequency of planet eccentricity for pre- and post-instability planets for the high-mass sample (runs 1-8) and the low-mass sample (run 9). Thicker gray lines represent Rayleigh distributions with scale factors 0.02 and 0.05. Right panel: median system eccentricity as a function of multiplicity, with the same colors as the left panel. Multiplicities are shifted slightly for clarity. Red crosses mark the power law (Equation 51) from \cite{He2020}.}
    \label{fig:ecc}
\end{figure}

A possible drawback of the instability model is the degree of dynamical heating from violent gravitational interactions. That is, orbit crossings entail growth in eccentricities and mutual inclinations \citep{Tremaine2015}. Figure \ref{fig:ecc} shows the distributions of orbital eccentricities in our simulations. Post-instability planets have eccentricities that approximately follow a Rayleigh distribution with scale parameter $\sigma_e$ that depends on the initial mass. For the high-mass sample, $\sigma_e\approx 0.05$, as has been seen in previous work \citep{Dawson2016,Izidoro2017}. For the low-mass sample, $\sigma_e\approx0.02$. Median system eccentricities are higher for higher intrinsic multiplicities, in line with the expectation from the maximum-AMD model of \cite{He2020}.

Eccentricity measurements of observed planets typically come from one of two methods. The first is a forward modeling approach that treats eccentricities and mutual inclinations as underlying distributions, along with other system parameters. Synthetic systems are then compared to observations; in particular, transit durations are the primary constraint on eccentricity \citep{Ford2008}. Such studies tend to recover scale parameters of $\sim 0.05$ \cite{Xie2016,VanEylen2019,Mills2019,He2020}. More recent work has suggested evidence for a multiplicity dependence on the distribution parameters, although that requires additional assumptions on system architecture as well as observational constraints from transit duration variations to confirm \citep{He2020,Millholland2021}.

The second method is through transit timing variations (TTVs), which, while typically used to measure masses, also depend strongly on eccentricity \citep{Agol2005a}. Statistical studies with this technique recover smaller eccentricities, with scale factor $\sim 0.02$ \citep{Hadden2014}. Furthermore, TTV systems have been shown to be significantly more circular than required for stability \citep{Yee2021}. However, two important biases limit the conclusions that can be drawn from TTV-derived eccentricities. Planet mass and eccentricity are degenerate in most cases, and hence eccentricity distributions depend on the choice of mass prior \citep{Hadden2017}. Additionally, TTV systems are a non-uniform sample of the multiplanet system population that preferentially selects planet pairs that are close to resonance and coplanar. In the instability scenario, these may be the systems that remained stable and did not experience growth in eccentricity. Or, if they did encounter an instability, they may have experienced a smaller degree of scattering that left them unusually coplanar and circular \citep{Esteves2020}.

For simplicity, our simulations were confined to the plane. In reality, planets are expected to exit the protoplanetary disk with inclinations of $\sim0.1^\circ$. To test the impact of small but non-zero inclinations, we repeated runs 1-5 starting from the mass loss step but gave each planet an inclination drawn uniformly from $[0^\circ,0.1^\circ]$. Because first-order resonances do not depend on inclination, the resonant angles continued to librate until the instability was triggered. The final intra-system uniformity in masses and period ratios was consistent with the results of the planar simulations. However, because in three dimensions orbital eccentricities can grow larger without guaranteeing a collision, collisional velocities were $\sim 20\%$ higher and the final eccentricity distribution had a longer tail past $e\sim0.1$. These results are consistent with the trends seen by \cite{Matsumoto2017}. Nevertheless, this set of inclined simulations likely overestimates eccentricities somewhat because the final planets have $\sim 30\%$ larger masses than in the coplanar simulations, which are themselves larger than typical super-Earth masses.

Even if eccentricities from post-instability systems are higher than those that are observed, this is not evidence against the instability model. Planet pairs just wide of mean-motion resonances require a mechanism that damps eccentricity after disk dissipation and this mechanism could operate in non-resonant systems as well \citep{Lithwick2012,Batygin2013}. Future work should determine to what extent tides or planetesimal scattering can reproduce the observed eccentricity distribution and whether such damping leaves observational signatures that can constrain post-nebular evolution.


Beyond consideration of the angular momentum deficit itself, collisions may effect a preference for ordering in systems by mass \citep{Ogihara2015}. Our initial conditions have no ordering as planet masses are chosen randomly. However, in real systems, planets tend to increase in mass and radius as orbital radius increases \citep{Millholland2017,Weiss2018}. To quantify any ordering in mass, we adapt the metric from \cite{Weiss2018} that considers the fraction $f$ of planet pairs in which the outer planet is more massive; unordered systems have $f=0.5$. As collisions proceed, mass tends to settle close to the star; by the end of our simulations, $40-50\%$ of planet pairs have a more massive outer planet. This prediction of the model does not match observed trends wherein $65\%$ of planet pairs have a larger radius outer planet \citep{Weiss2018} and a similar ordering exists in mass \citep{Millholland2017}. Planet radii measured from transit observations include atmospheres that may be strongly affected by photoevaporation or tidal heating \citep{Millholland2019a} and are therefore not a reliable estimate of mass \citep{Chen2017}. However, the presence of a marginally significant, but similar trend in mass measurements highlights a shortcoming of the instability scenario. A possible solution could be to consider a mass ordering in the initial conditions as an outcome of the planet formation process that is later partially eroded by collisions. Another potential process may be additional post-nebular accretion of left-over debris. These avenues for continued quantification of the instability mechanism as the process responsible for shaping the terminal architectures of exoplanet systems are worthy of investigation as their post-nebular evolution comes into sharper focus.

\acknowledgements
We are thankful to Erik Petigura, Juliette Becker, Fred Adams, Andrew Howard, and Lauren Weiss for insightful discussions. We are especially grateful to the anonymous referee, whose input significantly improved this work. K.B. is grateful to Caltech, the David and Lucile Packard Foundation, and the Alfred P. Sloan Foundation for their generous support.

\appendix
\section{Migration prescription}
\label{app}
Here we specify details of our ad-hoc migration prescription to construct the original resonant chains. 
The migration timescale is
\begin{equation}
    t_m = \frac{a}{\dot{a}} = -\frac{2\times 10^5}{\log_{10}{(r/\text{AU})} + 1} \text{ yr}
\end{equation}
where $r$ is the orbital radius. The timescale for eccentricity damping is
\begin{equation}
    t_e = \frac{e}{\dot{e}} = -\frac{2\times 10^2}{r/\text{AU}} \text{ yr}.
\end{equation}

The purpose of such a prescription is as follows. Capture into resonance depends only on the relative migration rate between a pair of planets. Denoting the inner planet by $1$ and the outer planet by $2$, that rate is
\begin{equation*}
    \frac{1}{t_{m,2}} - \frac{1}{t_{m,1}} = \frac{\dot{a}_2}{a_2} - \frac{\dot{a}_1}{a_1} = \frac{\log{r_1/r_2}}{2\times 10^5 \text{ yr}} = \frac{\log{P_1/P_2}}{3\times 10^5 \text{ yr}}.
\end{equation*}
For $P_2 > P_1$, this rate is negative, and migration is always convergent. Furthermore, the migration rate depends only on the period ratio and not the radius or period itself. The normalization constant in the denominator causes no migration at $0.1$ AU. The eccentricity damping timescale is chosen to be approximately two orders of magnitude smaller than the relative migration rate, in line with typical disk models \citep{Tanaka2004, Cresswell2008}.

\begin{figure}[h]
    \centering
    \includegraphics[width=0.8\textwidth]{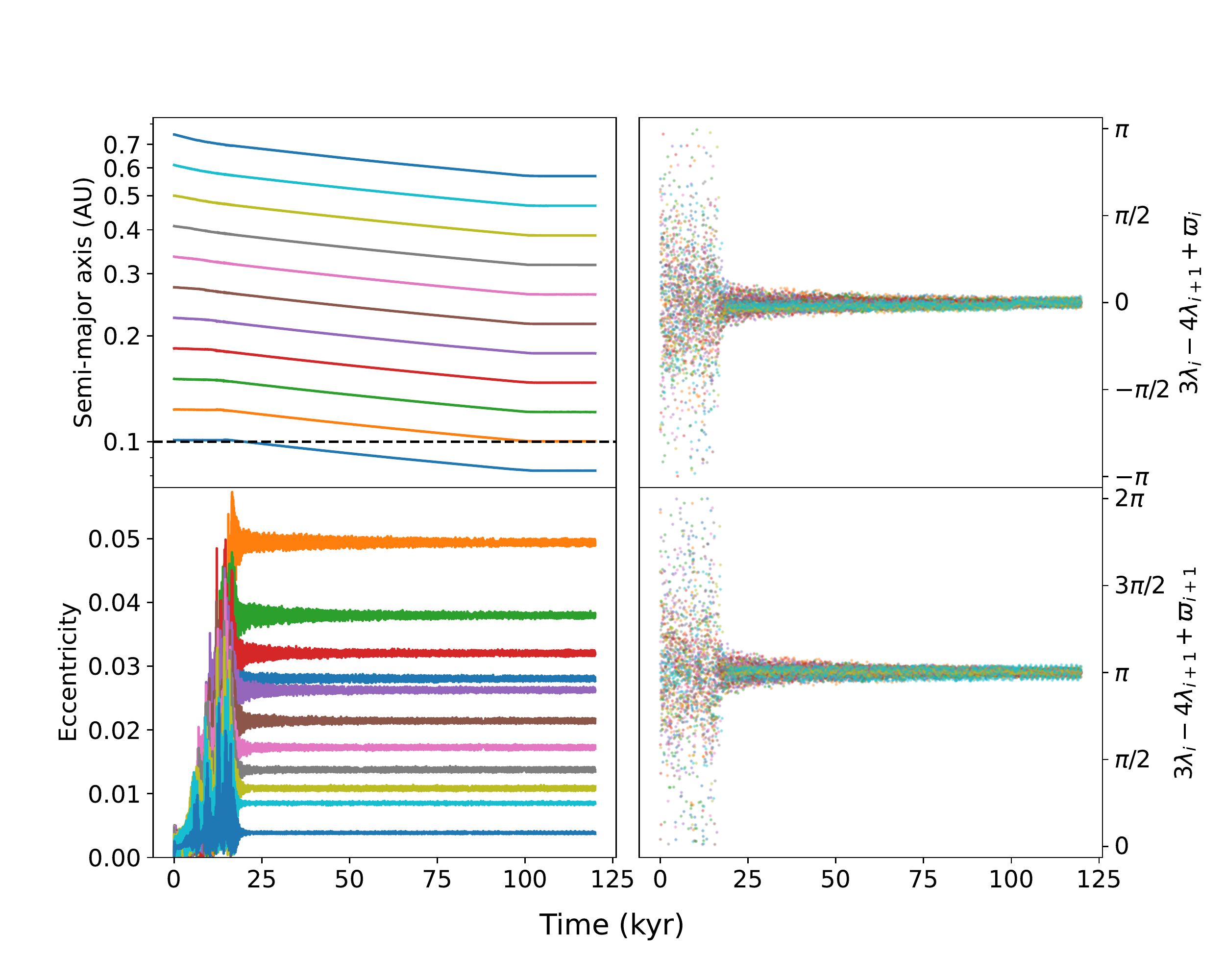}
    \caption{Evolution of orbital elements during a typical simulation of capture into a resonant chain. Here, the planet parameters are those of Run 1 (see Table \ref{tab:sims}). Migration and eccentricity damping proceeded from $t=0$ to $t=100$ kyr, at which point $t_m$ and $t_e$ increase exponentially, representing gas disk removal. At $t=110$ kyr, both timescales are set to infinity.}
    \label{fig:cap}
\end{figure}

Figure \ref{fig:cap} shows a typical capture into a resonant chain using this prescription. Planets spaced just wide of the intended resonance smoothly capture into the resonance and all 20 angles librate. The final eccentricities are consistent with more physically-motivated simulations \citep{Izidoro2017}.

\bibliography{main}
\bibliographystyle{aasjournal}

\end{document}